\documentclass[12pt,preprint]{aastex}
\usepackage{amsmath}

\shorttitle{PKS\,2152-699's X-ray Spectra}
\shortauthors{Ly, De Young, \& Bechtold}
\begin{document}
\title{The Discovery of Extended Thermal X-ray Emission from PKS\,2152-699: Evidence for a `Jet-cloud' Interaction}
\author{Chun Ly,\altaffilmark{1} David S. De Young,\altaffilmark{2} and Jill Bechtold\altaffilmark{3}}

\altaffiltext{1}{Present address: Steward Observatory, University of Arizona, 933 North Cherry Avenue,
                 Tucson, AZ 85721; cly@u.arizona.edu}
\altaffiltext{2}{National Optical Astronomy Observatory, 950 North Cherry Avenue, Tucson, AZ 85719;
		 ddeyoung@noao.edu}
\altaffiltext{3}{Steward Observatory, University of Arizona, 933 North Cherry Avenue, Tucson, AZ 85721;
                 jbechtold@as.arizona.edu}
\begin{abstract}
A \emph{Chandra} ACIS-S observation of PKS\,2152-699 reveals thermal emission from a diffuse region around
the core and a hotspot located 10\arcsec~northeast from the core. This is the first detection of thermal
X-ray radiation on kiloparsec scales from an extragalactic radio source. Two other hotspots located
47\arcsec~north-northeast and 26\arcsec~southwest from the core were also detected. Using a Raymond-Smith model,
the first hotspot can be characterized with a thermal plasma temperature of 2.6$\times10^6$~K and an electron
number density of 0.17~cm$^{-3}$. These values correspond to a cooling time of about 1.6$\times10^7$~yr. In
addition, an emission line from the hotspot, possibly \textsc{Fe xxv}, was detected at rest wavelength 10.04\AA.\\
\indent The thermal X-ray emission from the first hotspot is offset from the radio emission but is coincident
with optical filaments detected with broadband filters of \emph{HST}/WFPC2. The best explanation for the X-ray, radio,
and optical emission is that of a `jet-cloud' interaction.\\
\indent The diffuse emission around the nucleus of PKS\,2152-699 can be modeled as a thermal plasma with a
temperature of 1.2$\times10^7$~K and a luminosity of 1.8$\times10^{41}$~erg s$^{-1}$. This emission appears
to be asymmetric with a small extension toward Hotspot A, similar to a jet. An optical hotspot (EELR) is seen
less than an arcsecond away from this extension in the direction of the core. This indicates that the
extension may be caused by the jet interacting with an inner ISM cloud, but entrainment of hot gas is 
unavoidable. Future observations are discussed.
\end{abstract}

\keywords{galaxies: active --- galaxies: individual (PKS 2152-699) --- galaxies: jets --- galaxies: kinematics and dynamics}

\section{INTRODUCTION}
\label{1}
\indent Radio images have shown that jets of radio galaxies disperse on kiloparsec or megaparsec scales,
forming hotspots and lobes at the end of the jet's path. The formation of these lobes is believed to be caused
by the interaction between a jet propagating through the interstellar or the intergalactic medium
\citep[e.g.,][]{dave02}. In principle, the jet may be deflected by a dense inhomogeneity in the ambient medium,
and past three-dimensional numerical simulations have examined the physics of this interaction
\citep{dave91, higgins99, wang00}.\\
\indent When a jet interacts with the ambient medium, shocks are likely to form, producing extended emission
lines in the UV, optical, and X-rays through ionization of the medium \citep{tadhunter02}. Powerful radio
galaxies that may have undergone jet-cloud interactions that cause their extended emission-line regions (EELRs)
are 3C\,277.3 \citep{breugel85,tadhunter00,solorzano03}, 3C\,265 \citep{solorzano03}, PKS\,2250-41
\citep{villar99}, and PKS\,2152-699 \citep{tingay96, fosbury98}.\\
\indent PKS\,2152-699, a well-studied Fanaroff-Riley type II radio galaxy \citep{fanaroff74} at z = 0.0283, is
one of the brightest sources in the southern sky with a flux of $\sim$19~Jy at 2.7~GHz \citep{wall94}. In
addition, this source is quite isolated: the nearest galaxy is 250~kpc away and less than 50 galaxies are within
a 3~Mpc radius \citep{tadhunter88}. Previously, \cite{tadhunter88}, \cite{tingay96}, and \cite{fosbury98} have
suggested that the deflection of the northern jet is conceivably caused by a `jet-cloud' interaction. For
example, radio observations obtained from the Australia Telescope Compact Array (ATCA) reveal a hotspot
(hereafter Hotspot A) located about 10\arcsec~north-east from the core. The detection of the northern lobe and
hotspot (hereafter Hotspot B) indicates a 20\arcdeg~deflection in the jet \citep{tingay96,fosbury98}. In this
paper, the analysis of archival data from the \emph{Chandra} X-ray Observatory is presented as evidence for
a jet-cloud interaction occurring at Hotspot A and the extension near the core in PKS\,2152-699.\\
\indent In \S~\ref{2}, the X-ray observation will be presented and compared with previously obtained radio
and optical images. The analysis of X-ray spectra and the core's point-spread function (PSF) are provided in
\S~\ref{3}. In \S~\ref{4}, a jet-cloud interaction is presented to explain Hotspot A and the northeast, extended
thermal emission. \S~\ref{5} will describe possible future observations, and concluding remarks will be made in
the final section.

\section{OBSERVATIONS}
\label{2}
\subsection{Radio and Optical Observations}
\indent The ATCA was used by Fosbury et al. to observe PKS\,2152-699 at 3.5~cm (8.64~GHz) on 1992 January 19
and at 6.3~cm (4.74~GHz) on 1992 April 1. In addition, a \emph{HST}/WFPC2 observation (Proposal 5479, PI: M.
Malkan) published in \cite{fosbury98} was obtained with the F606W filter on 1994 October 20. The reduced radio
and optical images were provided by Fosbury et al.

\subsection{X-ray Observation}
\begin{figure}[!htc]
\epsscale {.60}
\plotone{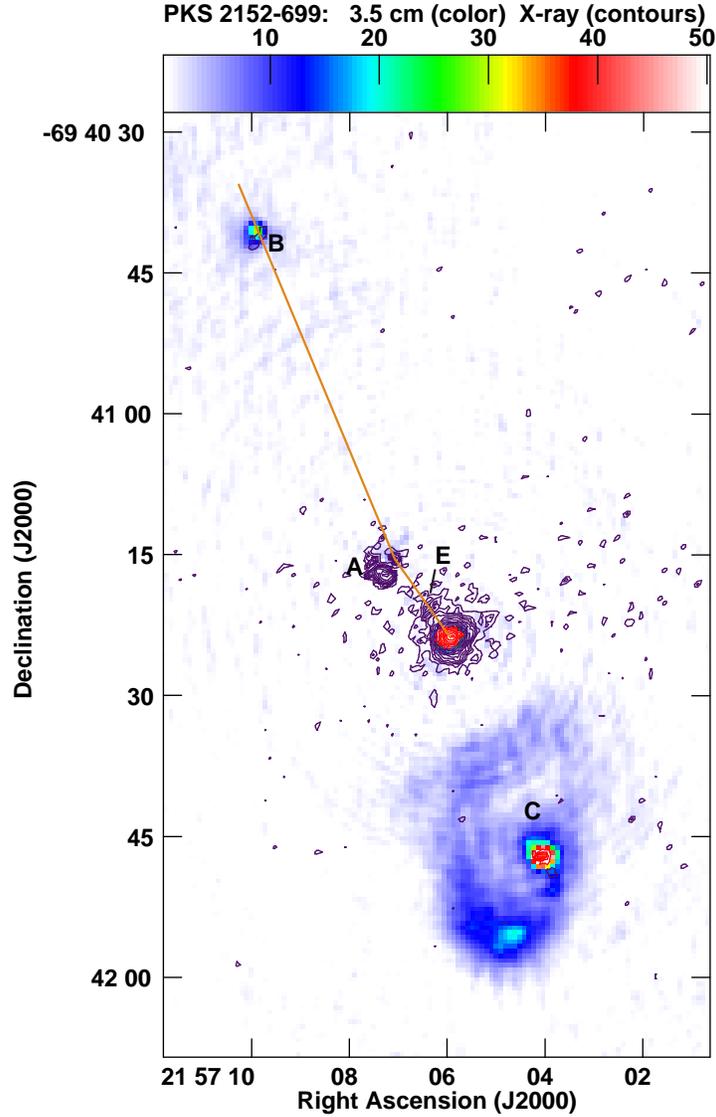}
\caption{The radio (3.5~cm) observation overlayed with X-ray contours of PKS\,2152-699. The red and purple
contours correspond respectively to regions where the radio flux density is greater than or less than 70
percent of the peak, 50~mJy beam$^{-1}$. Here, three hotspots are detected in both images, with an X-ray - radio
coincidence for Hotspot B and C. The low-level contours reveal that the diffuse emission is extended more
toward Hotspot A, and that it is also extended in the west-east direction. The colorscale levels range from
0.5 to 50~mJy beam$^{-1}$ with contours levels of 1, 2, 2.8, 4, 5.7 counts and multiples of $\sqrt{2}$
thereafter. The orange line represents the deflection of the northern jet at Hotspot A.}
\label{fig1}
\end{figure}

\begin{figure}[htp]
\plotone{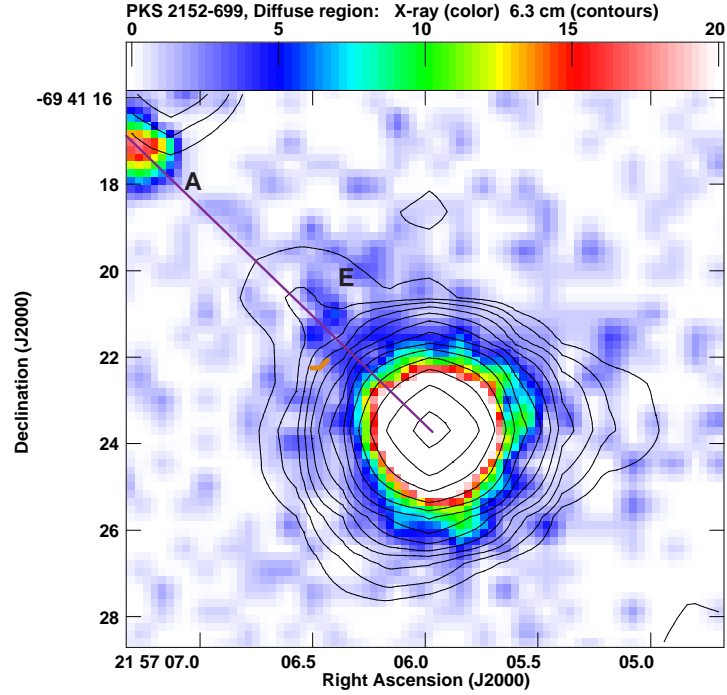}
\caption{Overlay between the X-ray image with 6.3~cm radio contours for the diffuse region. The contour levels
are 5~mJy beam$^{-1}\times$(2, 2.8, 4, 5.7, 8 and multiples of $\sqrt{2}$ thereafter) with the colorscale
ranging between 0 and 20 counts. The asymmetry in the diffuse emission (labeled as E) is coincident with the
radio jet. The purple line represents the radio jet axis, which is also parallel to the X-ray jet and passes
through Hotspot A's X-ray emission in the upper left. The orange line corresponds to the location of the optical
hotspot with its size indicated by the extent of the line.}
\label{fig2}
\end{figure}

\footnotetext[1]{\emph{Chandra} Proposers' Observatory Guide 2003: http://cxc.harvard.edu/proposer/POG/pdf/ACIS.pdf.}
\indent PKS\,2152-699 was observed (ObsID 1627, PI: A. Wilson) with the \emph{Chandra} Advanced CCD Imaging
Spectrometer (ACIS-S)\footnotemark[1] without a transmission grating in place on 2001 August 2 for
approximately 14~ks. The level 1 products were obtained directly from the archive where they went through
the standard data processing with the afterglow correction removed (See \S~\ref{psf}). In this observation, the
core with diffuse emission around it and three hotspots were detected, as shown by the contours in Figure~\ref{fig1}.
The most prominent hotspot is located 10\arcsec~(5.6~kpc, PA 137\arcdeg) from the core. One arc-second corresponds
to 0.56~kpc under the assumption that H$_{\circ}$ = 71~km~s$^{-1}$~Mpc$^{-1}$, $\Omega_{\lambda}$ = 0.73, and
$\Omega_m$ = 0.27 \citep{bennett03}. The other two hotspots, B and C, were detected
47\arcsec~(26.3~kpc, PA 116\arcdeg) and 26\arcsec~(14.6~kpc, PA 292\arcdeg) from the core.\\
\indent The X-ray contours in Figure~\ref{fig1} reveals that the diffuse emission is extended more toward the
northeast direction. This extension (labeled as E) is similar to a jet when it is compared to the 3.5~cm radio
jet as shown in Figure~\ref{fig2}. The purple line shows that the radio jet propagates toward Hotspot A's X-ray
emission. In addition, the diffuse emission extends in the east-west direction as noted by the low-level contours
in Figure~\ref{fig1}.

\subsection{Positional Corrections in Images}
Prior to overlaying these images, positional corrections relative to the International Celestial Reference Frame
Extension 1 (ICRF-Ext.1) coordinates were made \citep{ma98,iers99}. The offsets
($\sigma_{\alpha}$, $\sigma_{\delta}$) in arc-seconds between the ICRF-Ext.1 coordinates with the
\emph{Chandra} and \emph{HST} coordinates are (0.1, -0.1) and (-1.4, -0.1), respectively. The radio observations
had a correction of (-0.4, -0.3). The X-ray position agrees with surveys\footnotemark[2] indicating an offset
within 0.4\arcsec~relative to the ICRF-Ext.1, and the optical offsets are reasonable compared to the point
inaccuracy of 0.26 to 1.84\arcsec~\citep{biretta96}. These images were imported into the Astronomical Image
Processing System (AIPS) where the package HGEOM was used to geometrically align the images together.
\footnotetext[2]{In-Flight Calibration Observations: http://cxc.harvard.edu/cal/docs/cal\_present\_status.html.}

\subsection{Multi-wavelength Overlays}
\begin{figure}[htp]
\vspace{17cm}
\includegraphics{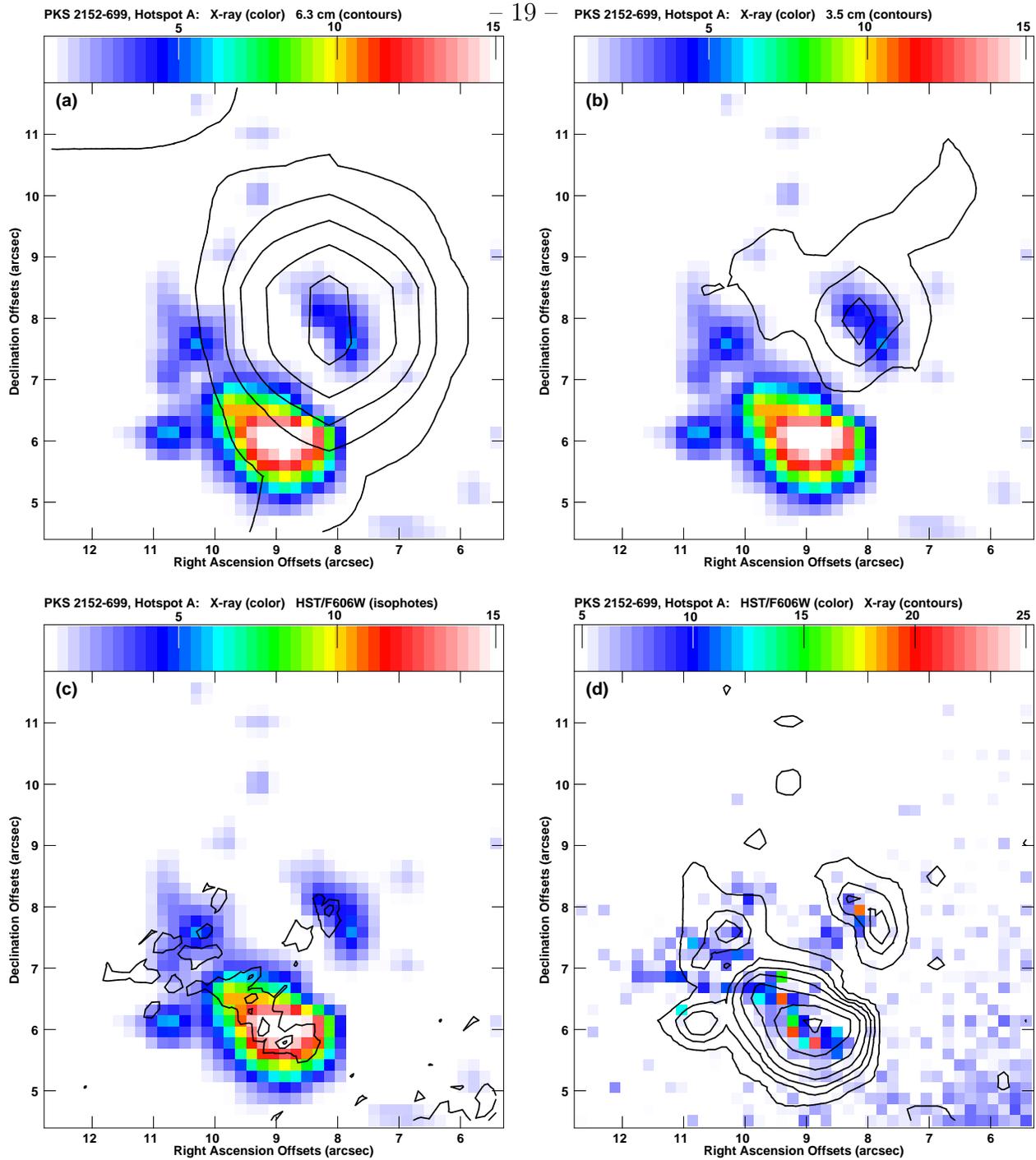}
{{}}
\caption{\tiny{Multi-wavelength images of Hotspot A in PKS\,2152-699. The contour levels are scaled from the lowest
level by 1, 2, 2.8, 4, 5.7, 8 and multiples of $\sqrt{2}$ thereafter.
({\em a}) An overlay between the 6.3~cm contours and the X-ray image: the lowest level is 5 mJy beam$^{-1}$;
the color-scale levels range from 0 to 15 counts.
({\em b}) An overlay between the 3.5~cm contours and the X-ray image: the lowest level is 2 mJy beam$^{-1}$;
the color-scale levels range from 0 to 15 counts.
({\em c}) An overlay between the \emph{HST}/F606W isophotes and the X-ray image: the lowest level is 7 counts;
the color-scale levels range from 0 to 15 counts.
({\em d}) An overlay between the X-ray contours and the \emph{HST} image: the lowest level is 1.5 counts;
the color-scale levels range from 5 to 25 counts. In ({\em a}) and ({\em b}), the radio hotspot is
2 to 3\arcsec~offset from the X-ray hotspot. ({\em c}) and ({\em d}) reveal an alignment between the
optical and X-ray emission for Hotspot A with some optical filaments not seen in the X-ray image.}}
\label{fig3}
\end{figure}

\indent In Figure~\ref{fig1}, an overlay between the X-ray image and 3.5~cm radio contours is shown. X-ray
emission from Hotspot B and C is coincident with those found in the radio images. Neither of these two hotspots
were within the \emph{HST}/WFPC2 field-of-view. The overlays for Hotspot A between the X-ray image
with the radio contours and the \emph{HST} isophotes are provided in Figures~\ref{fig3}a-c, while Figure~\ref{fig3}d
shows the X-ray contours on the \emph{HST} image. In both radio observations, Hotspot A is offset 2 to 3\arcsec~
from the X-ray hotspot, similar to what is seen for the case of the \emph{HST} image \citep{fosbury98}. A
comparison between the optical and the X-ray image reveals an alignment for Hotspot A; however, some filaments
visible in the \emph{HST} image are not seen in the X-ray image.\\
\indent A careful examination of the \emph{HST} image reveals an optical hotspot that is less than an arcsecond
away. The location is indicated by the orange line in Figure 2. 

\section{X-RAY ANALYSIS}
\label{3}
\indent The X-ray spectrum of Hotspot A, the diffuse region, and Extension E were analyzed using the
\emph{Chandra} Interactive Analysis of Observations (CIAO v3.0.2)\footnotemark[3]. For Hotspot B and C,
less than 40 counts were obtained, which is insufficient data to distinguish between thermal and non-thermal
emission processes. The source and background regions were selected, and the spectra of these regions were
extracted using the \textit{acisspec} script and imported into \textit{Sherpa} for modeling. Hotspot A's
extracted region is indicated in Figure~\ref{fig4}b by the green circular outline that has a radius of
3.5\arcsec. The diffuse emission's extracted region is identified by the green annulus that has an inner and
outer radius of 2.2\arcsec~and 7.6\arcsec, respectively. The cyan rectangular outline that is
4.5\arcsec$\times$3.4\arcsec~represents the extracted region of Extension E.\\
\indent To obtain an appropriate background that avoids the diffuse region and the hotspots, four circular
(10\arcsec~in radius) regions that are approximately positioned 39 to 44\arcsec~north-west, west, south, and
north-east from the core were selected. The count rates per square arcsecond are
3.2$\times10^{-5}$~cts s$^{-1}$ arcsec$^{-2}$ for the background regions,
2.4$\times10^{-4}$~cts s$^{-1}$ arcsec$^{-2}$ for the diffuse region, 5.2$\times10^{-4}$~cts s$^{-1}$ arcsec$^{-2}$ 
for Hotspot A, and 6.5$\times10^{-4}$~cts s$^{-1}$ arcsec$^{-2}$ for Extension E. Uncertainties reported later
in this paper for the models' parameters are 1\,$\sigma$ deviation, determined from the \textit{Sherpa} task,
\textit{projection}.
\footnotetext[3]{See http://cxc.harvard.edu/ciao/.}

\subsection{Hotspot A}
\indent Shown in Figure~\ref{fig4}a is Hotspot A's spectrum between 0.30 to 1.35~keV without redshift correction
but background subtracted. Initially, a non-thermal model (power-law with galactic absorption)
was tested, and the overall residuals indicated a poor fit. Plotted on the spectrum in blue is this non-thermal
model. The residuals are shown in the middle panel. The parameters for this power-law model are
$\Gamma$ = 2.6$\pm$0.4 with a normalization of (1.1$\pm$0.3)$\times10^{-5}$ and a galactic neutral
hydrogen column density\footnotemark[4] of 2.52$\times10^{20}$~cm$^{-2}$. The reduced chi-squared per
degrees of freedom is $\chi^2_{69}$ = 0.42.\\
\indent Further analysis reveals that the Raymond-Smith model for a thermally ionized plasma is appropriate
between 0.3 to 1.0~keV. The temperature for this model is 0.22$^{+0.04}_{-0.03}$~keV ($2.6\times10^6$~K) with
a metal abundance of [Z/Z$_{\Sun}$] = 0.10$_{-0.15}^{+8.24}$ and $\chi^2_{68}$ = 0.36. However, this model was
not able to explain a peak between 1.1 and 1.3~keV, so an additional Gaussian component was added to model an
emission line. Using the \textit{Sherpa} package GUIDE for line identification, this emission was identified as
\textsc{Fe xxv}, although the uncertainties in the peak are large enough for other possible emission lines
(e.g., Fe$^{19+}$, Ni$^{19+}$). The error in the model, shown below the spectrum, fluctuates randomly
between $\pm$1.5\,$\sigma$ with a $\chi^2_{65}$ = 0.33.\\
\indent The emission from Hotspot A is thermal based on the comparison between the residuals. Using the F-test
for a thermal model that excludes and includes the emission line, the significance is about 0.05, indicating
that the better model involves the emission line. Therefore, the best fit for the hotspot is
(\textsf{rs}+\textsf{gauss})\,$\times$\,\textsf{abs} where \textsf{rs} is the \textit{XSpec} Raymond-Smith
model (xsraymond), \textsf{gauss} is a simple one-dimensional normalized Gaussian model (ngauss1d) and
\textsf{abs} represents galactic absorption (xswabs). In Figure~\ref{fig4}a, the red line plotted on the spectrum
is the best model with the residuals in the lowest panel.\\
\footnotetext[4]{Obtained from Colden on the \emph{Chandra} Proposer page: http://cxc.harvard.edu/toolkit/colden.jsp.}

\subsection{Diffuse Region}
\indent The spectrum for the extended region is shown in Figure~\ref{fig4}c between 0.3 to 2.6~keV. Once
again, a power-law fit was first attempted. This fit was reasonable for the tail of the spectrum between
1.2 to 2.6~keV, but a poor fit elsewhere since the peak in the fit is off from the source spectrum, and the
residuals have a ``bump'' profile between 0.5 to 1.2~keV. The parameters for this non-thermal model are
$\Gamma$ = 1.7$\pm0.2$ with a normalization of (3.8$\pm$0.3)$\times10^{-5}$ and $\chi^2_{162}$ = 0.33.\\
\indent Our second attempt was to consider a thermal bremsstrahlung model (xsbremss). The profile for
this model was similar to the power-law case and hence also a poor fit. This model yielded a goodness of fit of
$\chi^2_{162}$ = 0.30. The temperature obtained from this fit is 1.4$_{-0.3}^{+0.5}$~keV (1.6$\times10^7$~K)
with C = 7.4$_{-1.0}^{+1.1}\times10^{-5}$. Here C = (3.02$\times10^{-15}/4\pi D^2)\int n_e n_H~dV$, where
$D$ is the distance to the source in cm, $n_e$ is the electron density in cm$^{-3}$, and $n_H$ is the hydrogen
density in cm$^{-3}$.\\
\indent To improve the spectral modeling, the Raymond-Smith model was considered. This fit modeled the
data better than the two previous attempts for energies between 0.8 to 1.2~keV, had the lowest
$\chi^2_{154}$, 0.28, and  the residuals appear flatter. The temperature is $1.04_{-0.15}^{+0.28}$~keV
($1.2\times10^7$~K) with a metal abundance of [Z/Z$_{\Sun}$] = 0.07$_{-0.07}^{+0.06}$ and a normalization C
of (2.1$\pm$0.4)$\times10^{-4}$. Although all three models look similar in the top plot of Figure~\ref{fig4}c,
the residuals of the thermal model indicate that the emission is thermal. Hence, \textsf{rs}\,$\times$\,\textsf{abs}
is the best model for the diffuse region.\\
\indent The Raymond-Smith variable model (xsvraymond) was considered, and the temperature was similar to the
latter attempt. However, it yielded certain metal abundances that were significantly greater than solar
abundances with large uncertainties. Although 450~counts were obtained, additional data are needed to better
constrain individual metallicities.\\
\indent Table~\ref{table1} summarizes the best model for Hotspot A and the diffuse region with the non-thermal
attempts. The only parameters held fixed during modeling are the known redshift (in the case of \textsf{rs})
and the galactic neutral hydrogen density. For the Gaussian model, only an upper limit on the FWHM exists, and
it is greater than the expected value.\\

\subsection{Extension E and PSF Modeling}
\label{psf}
\indent Extraction of the northeast extended emission within the diffuse region reveals similar thermal
properties as the entire diffuse region. Extension E has an ionized plasma temperature of 1.0$_{-0.5}^{+3.8}$~keV
($1.1\times10^7$~K) with an abundance [Z/Z$_{\Sun}$] of 0.09$_{-0.09}^{+0.70}$.\\
\indent Prior to extracting the PSF, the afterglow\footnotemark[5] events were examined, and were not removed in
the final level 2 processing because most of these events were from the core (an additional 170 counts). PSF
modeling (using mkpsf) of the core at 1.5~keV indicates that the diffuse emission is real compared to the wings
of the core. The wings of the normalized PSF fell below one count beyond 2\arcsec~from the core, which is
within the inner radius of the annulus in Figure~\ref{fig4}b.
\footnotetext[5]{See the ``Remove the acis\_detect\_afterglow Corrections'' thread at
http://cxc.harvard.edu/ciao/threads/acisdetectafterglow/ for more information.}

\subsection{Number Densities and Cooling Times}
\subsubsection{Hotspot A}
\indent The Raymond-Smith model for Hotspot A gives a total luminosity of 5.2$\times10^{40}$~erg s$^{-1}$
or 4.6$\times10^{40}$~erg s$^{-1}$ excluding the Gaussian model. Assuming that the thermal X-ray emission of
Hotspot A is produced via shock-induced processes (See \S~\ref{4.1.2}), the post-shock temperature derived from the
Rankine-Hugoniot conditions is
\begin{equation}
T = \frac{3\mu m_p v_{sh}^2}{16k} \Rightarrow v_{sh} = \left(\frac{16kT}{3\mu m_p}\right)^{1/2},
\end{equation}
where $v_{sh}$ is the shock velocity \citep{landau79}. For a temperature of 2.6$\times10^6$~K, the shock velocity is
$\sim$\,480 km s$^{-1}$. The kinetic energy flux from a shock is $\case{1}{2}\mu nm_p v_{sh}^3A_{sh}$, where $A_{sh}$
is the area of the shock region. Assuming that the X-ray luminosity is half the kinetic energy flux and that the shock
region consists of the bright area in the X-ray image (Figure~\ref{fig3}) of $\sim$\,1.3~kpc$^2$, then the number
density is 0.17~cm$^{-3}$.\\
\indent The hotspot's emission is projected over a rectangular region outlined in white in Figure~\ref{fig4}b.
For a projected emitting region of 4.4\arcsec\,$\times$\,4.7\arcsec~ (2.5~kpc\,$\times$\,2.6~kpc) and assuming
a depth of 2~kpc, the emitting volume is 13.0~kpc$^{3}$ and the power radiated per volume (L$_B$) by the
ionized plasma (excluding the emission line) is 1.2$\times10^{-25}$~erg s$^{-1}$~cm$^{-3}$. The cooling time
for the plasma can be determined from the ratio of the power radiated per volume and the energy density:
\begin{equation}
t_c = \bigg[\frac{L_B}{E}\bigg]^{-1} = \frac{n_e k T}{L_B}.
\label{eqn2}
\end{equation}
For the numbers reported above, the cooling time for the plasma is approximately 1.6$\times10^7$~yr.\\

\subsubsection{Diffuse Region}
With the plasma temperature and luminosity known for the region, the electron number density $n_e$ and the
cooling time $t_c$ for the diffuse region can be determined. The power radiated per unit volume $L_B$ for a hot
plasma is defined as
\begin{equation}
L_B = \frac{L}{V} = \Lambda(T) n_e^2,
\label{eqn1}
\end{equation}
where $\Lambda$($T$) is the total cooling coefficient in units of erg cm$^3$~s$^{-1}$, and $V$ is the volume
of the emitting region. For a temperature of $\sim$10$^7$~K, $\Lambda$ is approximately
 7$\times10^{-23}$~erg~cm$^3$~s$^{-1}$ \citep{raymond76}.\\
\indent For the diffuse emission without the extension, assuming that the volume is spherically symmetric and
that the emitting region is projected onto the sky as an annulus, the volume is 305~kpc$^3$. The luminosity
of the diffuse emission is 1.4$\times10^{41}$~erg s$^{-1}$. This implies a number density of 0.01~cm$^{-3}$
and a cooling time of 5.1$\times10^7$~yr.\\
\indent The greatest uncertainty in our calculations is in the size of the emitting region. The emitting
region used in our calculation is (6.7~kpc)$^3$. The proportionality between $n_e$ and $t_c$ with
the volume is
\begin{equation}
\begin{split}
n_e \propto&~V^{-1/2},~\textrm{and}\\
t_c \propto&~V^{1/2}.
\end{split}
\label{eqn3}
\end{equation}
Thus, a factor of two in the volume will only decrease $n_e$ and increase $t_c$ by a factor of $\sqrt{2}$.\\
\indent For the extended section of the diffuse emission (indicated by the cyan box), the emitting volume is
9.8~kpc$^3$ when assuming that the depth is 2~kpc, and the luminosity is 3.2$\times10^{40}$~erg s$^{-1}$.
The number density then for the extended region in the diffuse emission is 0.04~cm$^{-3}$ with a cooling time
of 1.8$\times10^7$~yr.

\subsubsection{Broad-band Flux Plot}
\indent Figure~\ref{fig5} reveals the broad-band flux (between radio to X-ray) for the north-western component
of Hotspot A as seen in Figure~\ref{fig3}. Radio, optical and X-ray emission overlap at this position.
An additional measurement at 5472\AA~was reported by \cite{serego88}. The upper limit infrared (IR) fluxes from
\emph{IRAS} observations were provided by
\cite{golombek88} at 12 and 100~$\mu$m. PKS\,2152-699 was detected at 25 and 60~$\mu$m, but are not included in
Figure~\ref{fig5} because the poor resolution of \emph{IRAS} could not distinguish possible IR emission of
Hotspot A from the core. The optical flux of the \emph{HST}/F606W observation is primarily [\textsc{O iii}],
[\textsc{N ii}], and H$\alpha$ emission lines \citep{fosbury98}.

\section{INTERPRETATION}
\label{4}
\subsection{Hotspot A}
\indent The thermally ionized volume of gas is located about 5.6~kpc from the nucleus of the parent galaxy
\citep{fosbury98}, and there are two possible explanations that can account for the presence of X-ray emission
at this distance. One explanation considers the possibility that hot material from the inner regions of the
interstellar medium (ISM) is advected out to this location by the radio jet, and the other is the jet-cloud
interaction mentioned above.

\subsubsection{Entrainment from Inner Regions}
\indent The gas could have been entrained and heated from the dense ISM in the inner regions of the galaxy
by the outflowing jet. This process is virtually unavoidable over a wide variety of physical parameters
\citep[e.g.,][]{dave02}. If entrainment is to explain the X-ray emission from Hotspot A, then the cooling
time of the gas must be approximately equal to the transit time of the entrained material so that the
X-ray emission is visible only at the end of the entraining region.\\
\indent However, the fact that no corresponding thermal X-ray emission is seen to the south requires special
arrangements to exist, presumably in the form of different conditions in the inner ISM for the north jet than
for its southern counterpart. In addition, in this picture the proximity of radio and X-ray emission at
Hotspot A must be regarded as fortuitous coincidence. These issues, together with the interpretation of the
diffuse X-ray emission near the nucleus in the direction of the radio jet (see \S~\ref{diffuse}), lead us
to believe that the most plausible interpretation of the X-ray and radio morphology at Hotspot A is that of a
jet-cloud interaction. 

\subsubsection{`Jet-cloud' Interaction}
\label{4.1.2}
\indent The thermal emission could arise from the interaction of the radio jet and a relatively dense cloud of
ambient gas located in the path of the oncoming jet. This latter possibility was suggested by \cite{fosbury98}
as a mechanism to explain both the optical emission lines in the inner cloud and the apparent misalignment
between the inner and outer northern radio hotspots. The mass of the cloud is 5.3$\times10^7$~M$_{\Sun}$, which
is reasonable for large clouds in the ISM \citep{young91}. The mass could be even larger and still be within
conventional limits for giant clouds. The collapse time for this cloud due to its own self-gravity is
1.6$\times10^8$~yr, an order of magnitude longer than the cooling time but much shorter than the age of the
parent galaxy.\\
\indent A jet-cloud interaction can also explain both the presence of radio emission co-located with
the thermal X-ray emission as well as account for the misalignment. A grazing encounter between collimated
supersonic flow and a dense object will give rise to internal shocks in the flow as it is deflected away from
the massive object. These shocks will cause enhancements of both the relativistic particle and magnetic
field densities, and it will result in a significant brightening of the synchrotron radiation of the jet,
giving rise to a radio hotspot.\\
\indent The interaction will also cause a shock to propagate into the cloud as it deflects the jet. This
shock-heated gas will cool rapidly in the high density cloud, emitting thermal X-rays and highly ionized
emission lines as seen in this \emph{Chandra} observation. This hot, post-shock gas will reside in the
cloud and be adjacent to, but not exactly coincident with the jet. Thus the offset between the radio and
X-ray emission is naturally explained by this mechanism.\\
\indent Furthermore, the geometry of this offset is also consistent with a 20\arcdeg~jet deflection. The
northernmost radio hotspot is offset to the west from a straight line between the nucleus and Hotspot A,
and its radio emission is also offset to the west from the X-ray emission. This is consistent with the
jet grazing the cloud on the west side, deflecting the jet to the west of the original radio-jet axis
(see Figure~\ref{fig1}), and causing a shock to propagate into the interior of the cloud, most of which
lies east of the contact point (see Figure~\ref{fig3}).\\
\indent In addition, the shallow grazing incidence of the jet upon the cloud will produce a strong shock
propagating into the cloud. This interaction is consistent with the temperature derived here for the thermal
emission: T~$\sim 10^{6}$ K as a post-shock temperature implies shock speeds of a few hundred km~s$^{-1}$,
which are very much less than the probable jet speed of over 1000 km~s$^{-1}$.\\
\indent Numerical simulations of the interaction of the radio jets with dense clouds provide support for the
latter picture, though the parameters of these calculations may not fit the exact values required for the
case of PKS\,2152-699. For example, \cite{wang00} consider interactions with very dense clouds, hundreds to
thousands of times the jet density, and find that jets tend to be destroyed or disrupted by very massive
clouds. Longer term deflections occur if the cloud-jet density ratios are less extreme and the Mach numbers
are not large \citep{dave91,higgins99}, though in some cases the cloud can be destroyed by the jet if the
cloud is sufficiently underdense or of small mass. Clouds of $\sim10^7$~M$_{\Sun}$ with densities of order
100 times the jet density can survive over $10^7$ years when deflecting jets with speeds in excess of
$10^3$~km s$^{-1}$. Therefore, long-term jet deflection, especially for the small apparent deflection angles
seen here, can probably be produced from a combinations of reasonable jet speeds, jet-cloud density ratios,
and total cloud masses.\\
\indent Although both the entrainment and deflection pictures are physically plausible, the north-south
asymmetry of the radio jets and the presence of thermal X-ray emission suggest that the simplest
interpretation may well be that the northern jet of PKS\,2152-699 is slightly deflected by a massive
interstellar cloud, and this deflection is producing shock-heated gas in the cloud, which manifests itself
via the thermal X-ray emission reported in this paper.\\

\subsection{Diffuse Emission}
\label{diffuse}
\indent As discussed in the observational section, there is a faint X-ray extension of the diffuse emission
in the direction of the radio jet. This emission appears to be thermal rather than non-thermal,
and the temperature and metal abundance for this extension is the same as that of the rest of the diffuse
emission. The number density in this extension is roughly 0.04~cm$^{-3}$, which is four times greater than
the diffuse emission as a whole. It appears that this X-ray emitting material in the extension could well
be the same hot gas as that around the galaxy that is giving rise to the overall diffuse X-ray emission.\\
\indent A nearby optical hotspot is seen, and it may be that the extension is due to the same process as in
Hotspot A of a jet interacting with a dense cloud. However, since extended thermal emission (representable as
hot gas) is present next to the core, it is likely that the gas may also be extrained by the jet, tracing out
an extension.\\
\indent To summarize, the radio morphology of the northern jet and the radio - X-ray misalignment for Hotspot A
would indicate a jet-cloud interaction. Moreover, thermal emission for Hotspot A is explained by shocks formed
in the clouds when this interaction occurs. This interaction may also be occurring at Extension E where the
optical image reveals an EELR.

\section{FUTURE OBSERVATIONS}
\label{5}
\indent This 14~ks observation of PKS\,2152-699 detected three hotspots and diffuse emission around the core.
Although this observation was sufficient to confirm thermal emission from Hotspot A, a longer observation is
needed to constrain its metallicity and to determine the ambiguous emission line at approximately 10\AA.
Moreover, additional time will help distinguish whether thermal or non-thermal emission is detected in the
two farthest hotspots, and it will help understand the content of the inner ISM from improved modeling of the
diffuse region. Additional \emph{HST} observations of Hotspot B and C are needed since the optical flux will
constrain the mechanisms for the detected X-ray emission.\\
\indent Other sources that are believed to have undergone jet-cloud interactions should be investigated for
their X-ray emission properties. For example, radio and optical observations of Coma A (3C\,277.3) have
indicated a jet-cloud interaction from the coincidence of the emission-line gas and its double-radio lobes
\citep{breugel85}. There is also strong support for jet-cloud interactions for high redshift (z $>$ 0.6) radio
galaxies \citep[and references therein]{tadhunter02}. However, these sources have larger redshifts compared to
PKS\,2152-699, hence thermal emission may not be detected because these sources are more distant, and the
inverse-Comptonization of the CMB photons is more prominent.

\section{CONCLUSION}
\label{6}
\indent From previous X-ray observations of extended emission from extragalactic radio sources, only non-thermal
processes, such as inverse-Compton, Synchrotron Self-Compton, and Synchrotron radiation, are believed to be the
cause of X-ray (and possibly optical) emission from kiloparsec or megaparsec hotspots and jets
\citep{harris02,wilson03}. However, the X-ray observation of PKS\,2152-699 contradicts accepted ideas about
X-ray emission mechanisms. This observation reveals that thermal processes are dominant for Hotspot A and a
diffuse region around the core with an extension (much like a jet). Spectral modeling of the
hotspot reveals a Raymond-Smith plasma of 2.6$\times10^6$~K with an additional iron (or possibly nickel)
emission line. The diffuse region can also be modeled as a hot plasma of 1.2$\times10^7$~K.\\
\indent To explain the detection of thermal emission for Hotspot A, it is believed that a `jet-cloud' interaction
is occurring at the site of Hotspot A. The deflection of the northern jet and hotspot and the offset
between radio and X-ray emission for Hotspot A are evidence to support such an interaction. Moreover, as the
jet collides with the cloud, shocks are likely to propagate into the cloud, thus heating it. As the cloud cools
rapidly, thermal X-rays, as reported in this paper, are emitted.\\
\indent A comparison between the radio and X-ray images for the diffuse region reveals that it is extended
in the northeast direction with similar thermal properties as the diffuse emission. With a detected EELR
near Extension E, one possible explanation is that the jet is interacting with the cloud. However, the possibility
that hot gas (as seen by the diffuse emission) from the inner ISM is entrained by the jet can occur and will
contribute to the X-ray emission.\\
\indent This is the first detection of extended thermal emission on extragalactic scales from a diffuse
region and hotspot. Other radio sources believed to have undergone `jet-cloud' interaction should be
investigated for possible thermal X-ray emission.

\acknowledgments C. Ly appreciates Sally A. Laurent-Muehleisen and Eric Greisen for assistance in troubleshooting
problems with AIPS. We also thank an anonymous referee for suggestions that improved this paper. The National
Optical Astronomy Observatory (NOAO) is operated by the Association of Universities for Research in Astronomy
(AURA), Inc. under cooperative agreement with the National Science Foundation. The Australian Telescope is funded
by the Commonwealth of Australia for operation as a National Facility managed by CSIRO. This work is partly
supported by the NASA Arizona Space Grant Program and NOAO.

\begin{deluxetable}{llccccc}
\tablenum{1}
\tablecolumns{5}
\tablewidth{0pc}
\tablecaption{Parameters for the \textit{Sherpa} Models}
\tablehead{
& \colhead{(1)}      & \colhead{(2)}        & \colhead{(3)}         & \colhead{(4)}  \\
& \colhead{Parameter}& \colhead{Value}& \colhead{Lower bound} & \colhead{Upper bound}}
\startdata
\multicolumn{5}{c}{\textbf{Hotspot A}}\\
\multicolumn{4}{l}{(\textsf{rs}+\textsf{gauss})\,$\times$\,\textsf{abs}} & $\chi_{65}^2$ = 0.33\\
 \multicolumn{5}{l}{Raymond-Smith model (\textsf{rs})}\\
 &T (keV)       & 0.22 & 0.03 & 0.04\\
 &[Z/Z$_{\Sun}$]& 0.10 & 0.15 & 8.24\\
 &C             & 1.1$\times10^{-4}$ & 1.0$\times10^{-4}$ & 0.9$\times10^{-4}$\\
\multicolumn{5}{l}{Gaussian model (\textsf{gauss})}\\
 &$E_o$ (keV)   & 1.20 & 0.05   & 0.04\\
 &FWHM          & 0.005 & \ldots & 0.31\\
 &A             & 1.6$\times10^{-6}$ & 1.0$\times10^{-6}$ & 1.1$\times10^{-6}$\\\tableline
\multicolumn{4}{l}{\textsf{plaw}\,$\times$\,\textsf{abs}} & $\chi_{69}^2$ = 0.42\\
\multicolumn{5}{l}{Power-law model (\textsf{plaw})}\\
 &$\Gamma$      & 2.6   & 0.4    & 0.4\\
 &B             & 1.1$\times10^{-5}$ & 0.3$\times10^{-5}$ & 0.3$\times10^{-5}$\\\tableline\tableline
\multicolumn{5}{c}{\textbf{Diffuse region}}\\
\multicolumn{4}{l}{\textsf{rs}\,$\times$\,\textsf{abs}} & $\chi_{154}^2$ = 0.28\\
\multicolumn{5}{l}{Raymond-Smith model (\textsf{rs})}\\
 &T (keV)       & 1.04 &  0.15 & 0.28\\
 &[Z/Z$_{\Sun}$]& 0.07 &  0.07 & 0.06\\
 &C             & 2.1$\times10^{-4}$ & 0.4$\times10^{-4}$ & 0.4$\times10^{-4}$\\\tableline
\multicolumn{4}{l}{\textsf{plaw}\,$\times$\,\textsf{abs}} &$\chi_{162}^2$ = 0.33\\
\multicolumn{5}{l}{Power-law model (\textsf{plaw})}\\
 &$\Gamma$      & 1.7   & 0.2    & 0.2\\
 &B             & 3.8$\times10^{-5}$ & 0.3$\times10^{-5}$ & 0.3$\times10^{-5}$
\enddata
\label{table1}
\tablecomments{Parameters used in the \textit{Sherpa} models for Hotspot A and the diffuse region. 
Column 1 contains the parameters for the model, Column 2 indicates the corresponding values, and
Column 3 and 4 report the lower and upper 1\,$\sigma$ uncertainties. \textsf{rs} is the Raymond-Smith
model, \textsf{plaw} is a power-law model, and \textsf{gauss} is the one-dimensional Gaussian model. In
Column 1, T is the plasma temperature, Z is the metallicity with respect to solar abundances,
C = (3.02$\times10^{-15}/4\pi D^2)\int n_e n_H~dV$, $E_o$ is the center of the Gaussian peak, FWHM is the
full-width at half maximum of the Gaussian peak, A is the amplitude of the Gaussian normalization, and
B is the amplitude of the power-law model. The $\chi_{\nu}^2$ for each modeling attempt is reported.}
\end{deluxetable}

%\twocolumn

\newpage

\newpage

\newpage
\onecolumn

\newpage
\begin{rotate}

\begin{figure}[!htc]
\epsscale{.90}
\includegraphics[height=2.9335365in,width=7.5in,bb=5 505 825 788]{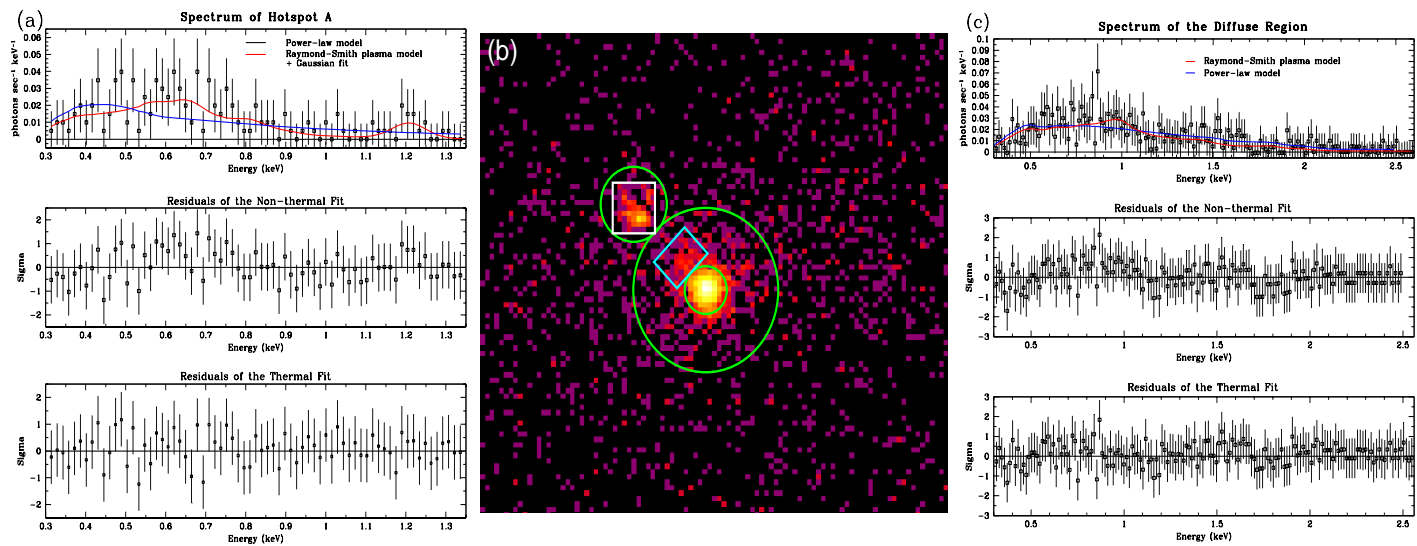}
\caption{The X-ray spectra of Hotspot A and the diffuse region. The left figure ({\em a}) shows Hotspot A's
spectrum obtained between 0.30 to 1.35~keV. The blue line represents the power-law fit, and the red line
represents the best fit, a Raymond-Smith model with a Gaussian centered at 1.2~keV. This emission line may
correspond to \textsc{Fe xxv}. Below the spectrum are the sigma residuals of the fits between 0.30 to 1.35~keV.
The extracted regions, outlined in green, are shown in the middle figure ({\em b}), and the emitting volume for
Hotspot A projected on the sky is shown by the white box. The cyan box indicates the region where Extension E's
spectrum was obtained to compare with the diffuse region. It is also the projected emitting region. The right
figure ({\em c}) shows the diffuse region's spectrum between 0.3 to 2.6~keV with the residuals of the fits below
it. The red and blue lines correspond to the Raymond-Smith and the power-law models, respectively.}
\label{fig4}
\end{figure}
\end{rotate}

\newpage
%\twocolumn

\begin{figure}[htp]
\plotone{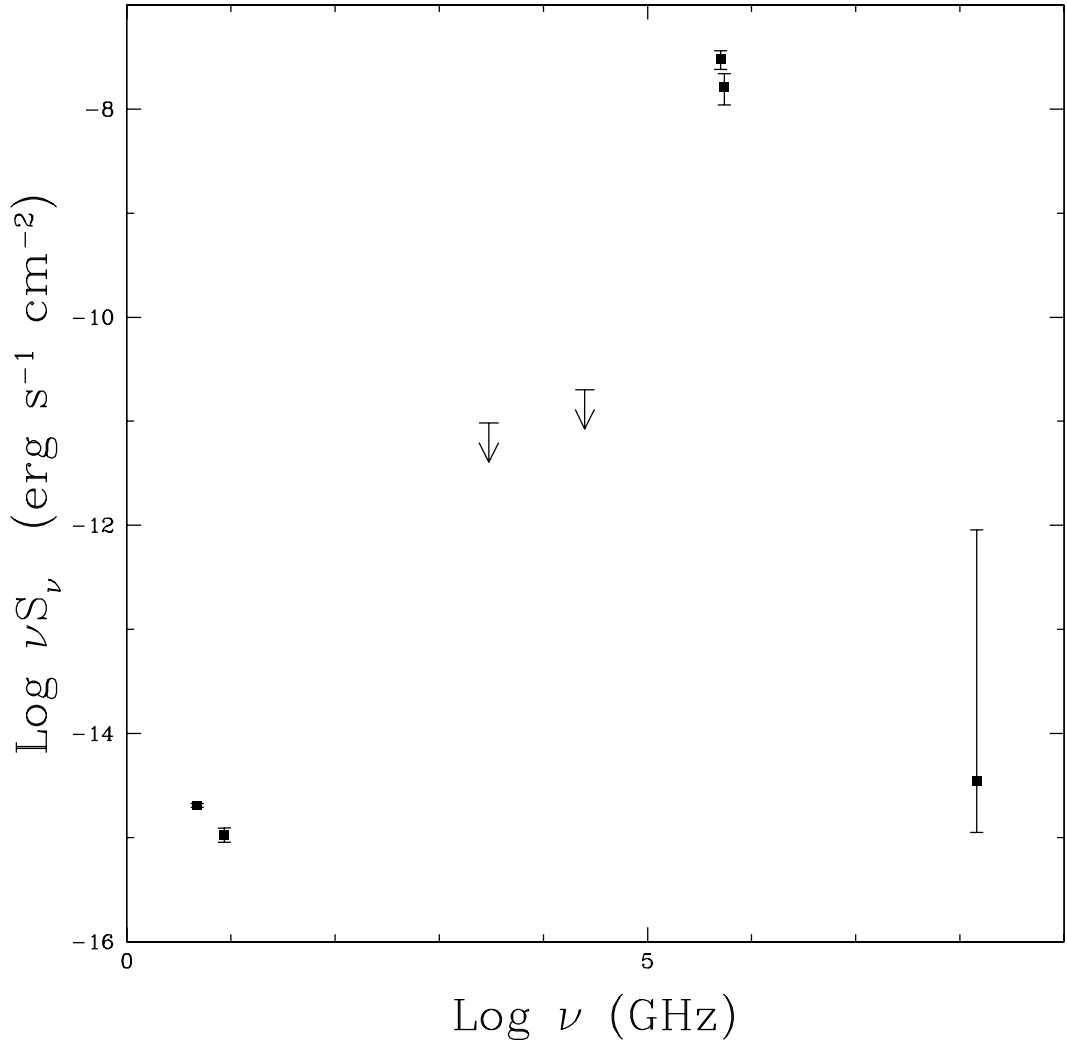}
\caption{A log-log plot of the broad-band flux $\nu\,S_{\nu}$ (erg s$^{-1}$ cm$^{-2}$) as a function of frequency
$\nu$ (GHz) for the unresolved north-western component of Hotspot A. The upper limits were provided by
\cite{golombek88} at 12 and 100~$\mu$m, and \cite{serego88} provided a measurement at 5472\AA.}
\label{fig5}
\end{figure}

\end{document}